# New features of CitedReferencesExplorer (CRExplorer)

Andreas Thor*, Werner Marx**, Loet Leydesdorff***, and Lutz Bornmann****

*University of Applied Sciences for Telecommunications Leipzig
Gustav-Freytag-Str. 43-45,
04277 Leipzig, Germany.
Email: thor@hft-leipzig.de

**Max Planck Institute for Solid State Research
Information Service
Heisenbergstrasse 1,
70506 Stuttgart, Germany.
Email: w.marx@fkf.mpg.de

***Amsterdam School of Communication Research (ASCoR),
University of Amsterdam,
P.O. Box 15793
1001 NG Amsterdam, The Netherlands.
Email: loet@leydesdorff.net

**** Corresponding author:
Division for Science and Innovation Studies
Administrative Headquarters of the Max Planck Society
Hofgartenstr. 8,
80539 Munich, Germany.
Email: bornmann@gv.mpg.de

Recently, we introduced the CitedReferencesExplorer (CRExplorer at http://www.crexplorer.net) (Thor, Marx, Leydesdorff, & Bornmann, 2016). The program was primarily developed to identify those publications in a field, a topic or by a researcher which have been frequently cited. It is especially suitable to study the historical roots of this field, topic or researcher (e.g., Barth, Marx, Bornmann, & Mutz, 2014; Marx, Bornmann, Barth, & Leydesdorff, 2014). CRExplorer analyses the cited references (CRs) of publications which can be downloaded from Web of Science (WoS, Thomson Reuters). In order to analyze the key publications with the CRExplorer, the user can use (1) a graph for identifying most frequently cited reference publication years (RPYs) and (2) the list of CRs which have been most frequently cited in specific RPYs.

Since the cited references are not fully standardized by the providers of literature databases, the program additionally provides a routine for the disambiguation of CRs. Several publications (e.g., Moed, 2005; Olensky, Schmidt, & van Eck, 2015) in bibliometrics have pointed to the problem of CR data that there exist variants of the same CR. In the new release of the CRExplorer, which has been published recently (see at http://www.crexplorer.net) the disambiguation feature has been significantly extended so that the program can also be used as a data preparation tool. Furthermore, not only data from Wos can be read and edited by CRExplorer, but also from Scopus (Elsevier). The data (from WoS or Scopus) can be exported in WoS or Scopus download formats. The export files can be imported in other programs for further processing, e.g. VOSviewer (van Eck & Waltman, 2010) or RPYS i/o (Comins & Leydesdorff, 2016).

Using the menu item "File" – "Import" – "Web of Science", CRExplorer opens one or more datasets from WoS (each download from WoS can contain up to 500 records). In WoS, the datasets are to be downloaded using the option "Save to Other File Formats". As "Record Content" select "Full Record and Cited References" and as "File Format" select "Other Reference Software". Using "File" – "Import" – "Scopus", however, CRExplorer can now also be used to read files from Scopus (Elsevier). The file format "CSV" (including citations, abstracts and references) should be chosen in Scopus for downloading records. Since the new release of CRExplorer is also enabled to export records, imports from Scopus can be exported as WoS files and imports from WoS can be exported as Scopus files. Thus, the program can be used as a tool to transfer the data from one format into another. Note that the export can only contain data from the import, when one transfers from Scopus to WoS. A WoS file, for example, does not consider more than a single cited reference (first) author while Scopus files include all authors in the CR field. If one transfers from WoS to Scopus, not all information can be provided by the input (e.g., the titles of referenced publications), since this information is not available in downloads from WoS.

CRExplorer can not only be used to transfer data between WoS and Scopus formats, but also to edit this data in an in-between step. For example, by selecting "Data" – "Remove by Cited Reference Year," the user can remove the data for specific RPYs. Using the menu items "Standardization" – "Cluster equivalent Cited References" and "Merge Cited References of the Same Cluster" the user can identify variants of the same CR, cluster them, and merge their occurrences (number of CRs). The clustering and merging of the data is especially important for the Scopus data, since the cited reference data is more heterogeneous than in WoS. Scopus data contains more information than WoS data (all authors and the titles of the referenced publication) which increases the risk of variants of the same CR. Furthermore, Scopus data may contain fragmented cited reference data which cannot be completely parsed into the bibliographic categories of CRExplorer (e.g. authors, titles, or volume numbers). Examples are as follows: "Rothschild: Where's the debate? (1971) New Scient, , (10 December), RS/ARF.879", "(1981) Reason, Truth, and History, 113p. , Cambridge", or "(2000) National Development Plan 2000 – 2006, , The Stationery Office, Dublin: 2000". The fragmented data of Scopus can be inspected at best by sorting the list of CRs under the column "Authors" in CRExplorer. A possible way of dealing with the fragmented CRs is to try their clustering and merging with complete CRs (if the fragmented CRs are variants of complete CRs). After finishing of the data editing in the CRExplorer, revised datasets can be exported in WoS or Scopus data formats.

We will continuously extend the features of CRExplorer. In the new release, for example, the user can select a specific CR in the CR table, press the space bar, and all bibliographic details of the CR are shown. Furthermore, we introduce a new internal file format "*.cre", which can be used as a "working format" and for the exchange of working files with colleagues. The file includes all data including matching results and manual matching corrections. Using Windows, one can double click on any *.cre file and thus run CRExplorer automatically. These and other new features are continuously updated in the description at http://www.crexplorer.net.

Since the transformation from the WoS to the Scopus format (and *vice versa*) and the disambiguation of CRs in-between are important tools for the preparation of data, which can then also be used in other bibliometric programs, we decided to signal the new features in a Letter to the Editor.